# Critical Regression Analysis of Real Time Industrial Web Data Set using Data Mining Tool

Shruti Kohli

Assistant Professor, Department of Computer Science and Engineering
Birla Institute of Technology, Mesra, Ranchi (India)

Ankit Gupta

Research Scholar, Department of Computer Science and Engineering
Birla Institute of Technology, Mesra, Ranchi (India),

## ABSTRACT
In today's fast pacing, highly competiting,volatile and challenging world, companies highly rely on data analysis obtained from both offline as well as online way to make their future strategy, to sustain in the market. This paper reviews the regression technique analysis on a real time web data to analyse different attributes of interest and to predict possible growth factors for the company, so as to enable the company to make possible strategic decisions for the growth of the company.

## 1. INTRODUCTION

Internet has become a platform where companies are growing more faster than the conventional market. While conventional market has its own pros and cons, internet has undoubtedly established itself as a main frontier's of future's global market. Last decade and specially last 5 years has seen a tremendous growth in the number of companies going online for retail business as well as providing different types of customer service in case of non retail companies. There is a cut throat competition among companies for (i) first acquiring the market share and then(ii) retaining their prospective customers. All the companies are striving very hard to remain in the competition and they are using all the means to collect and analyze data related to the companies so that they can prepare themselves for the future. They are leaving no space untouched to gather and analyze the required data, as this is the data which on correct analysis is supposed to take theses companies forward, while wrong interpretation of the data can be fatal for the companies also.

Researchers had been extensively using WEKA for their research activities. As reported by[1] Weka is a landmark system in the history of the data mining[7,12] and machine learning research communities because its modular, extensible architecture allows sophisticated data mining processes to be built up from the wide collection of base learning algorithms and tools provided. In their paper [1] sapna jain et al have provided a comprehensive details on evolution of WEKA and release,of k-means clustering execution in WEKA 3.7.

Another research work[2,14] shows the comparison of the different clustering algorithms and finds out which algorithm will be most suitable for the users. Most of the researchers had been using WEKA for data mining activites[1,2,3,4,5,11]. Data mining is the use of automated data analysis techniques to uncover previously undetected relationships among data items. Data mining often involves the analysis of data stored in a data warehouse. Three of the major data mining techniques are regression, classification and clustering.

However most of the researchers have not worked on conducting Regression technique analysis using Weka. Futher there are not many researchers that are mining the web data using WEKA. Although Weka does not include any "magical" web mining algorithms ("magical" meaning "one-click-fits-all").

However, the web usage information is either structured or semi-structured data[13] (which can easily be turned into fully structured). One can apply a lot of the algorithms weka offers. After deciding the kind of information one needs to extract from the web data one can use the algorithms and tools that are available with WEKA.

Endeavor of this paper is to high light the usage of WEKA in web mining. For any website it is very important to mine the web logs to determine whether the website is working fine or not. It also helps them to make future business strategies where they can decide the upon possible campaigns and over all budget required for the maintenance and growth of the website. Also it helps to forecast the future growth of the website under the current scenarios.

The research conducted in this paper uses a real time industrial dataset (table 1)(identity of the website has been masked) which was obtained from a mid sized company using Google Analytics.

This paper attempts to answer the following queries that are generated from our previous discussions:

1-Can WEKA be used as a tool to get future directions for a website?
2-Are WEKA Regression Techniques appropriate approach to take strategic future decisions. to help the company to remain in the market for in the long run.

Rest of this paper is organized as follows: Section 2 will be about Weka and Web Mining, Section 3 will be about experimental setup, section 4 will be the result analysis and section 5 will conclude this paper followed by references.

**Table 1. Data Set Details**

| Month  | ST    | BAS     | PPC     | REM   | VU  | PV     |
|--------|-------|---------|---------|-------|-----|--------|
| Apr-11 | 1k    | 0       | $650    | 0     | 100 | 20k    |
| May-11 | 5k    | 0       | $2,500  | 0     | 141 | 100k   |
| Jun-11 | 10k   | 0       | $3,000  | 0     | 90  | 200k   |
| Jul-11 | 12.5k | 0       | $1,600  | 896   | 50  | 250k   |
| Aug-11 | 15k   | 0       | $1,550  | 1,002 | 172 | 300k   |
| Sep-11 | 21k   | $6,234  | $3,500  | 1,788 | 300 | 400k   |
| Oct-11 | 31k   | $11,232 | $5,600  | 2,381 | 310 | 550k   |
| Nov-11 | 38k   | $7,439  | $3,900  | 2,979 | 310 | 650k   |
| Dec-11 | 40k   | $2,389  | $1,100  | 2,987 | 0   | 700k   |
| Jan-12 | 47k   | $7,823  | $3,900  | 2,916 | 54  | 800k   |
| Feb-12 | 56k   | $9,372  | $5,000  | 3,272 | 410 | 900k   |
| Mar-12 | 75k   | $18,782 | $10,500 | 3,987 | 410 | 1,150k |
| Apr-12 | 95k   | $18,378 | $11,000 | 4,876 | 415 | 1,300k |
| May-12 | 109k  | $17,283 | $7,500  | 5,498 | 500 | 1,475k |
| Jun-12 | 145k  | $35,986 | $19,500 | 5,007 | 510 | 1,900k |

## 2. USAGES OF WEKA FOR WEB MINING

Fig[1] provides a pictorial representation of the usage of Weka for web mining activities. As shown in the figure the data is extracted from the web analytical tools (here google analytics has been used) which act as the input for the WEKA tool. The tool provides comprehensive analytical reports based on the regression techniques which can be easily understood by the website owner for taking business decisions[10]. The web analytic reports are informative enough to provide the various dimensions on which company had been working and the quantitative value of the goal which company needs to achieve.

The company provided following information which has been used to identify the various variable and fixed components of the data.

1. Subscription to the website is very important, subscription to website is free of cost and company earns it's major revenue by selling CPM

2. PPC and banners are the major drivers of traffic and an important acquisition channel of the company

3. The company also sends monthly emails to remind existing members to come and visit the website to find out what's new offerings etc.

Using the above assumptions and the data set provided by the company following were the input and output variables determined for the current system

From figure 2, it can be visualized that no. of .subscriber's, Banner Ad Spend, PPC spend, Reminder Emails sent and Videos Upload are the various ways through the website tries to achieve the goal of improving page views. From these inputs it can be determined that efforts employed by the website can be categorized as follows:

Effort Based on Time: Company sends reminder emails to its subscriber base and uploads new videos on the website to increase visits and page views

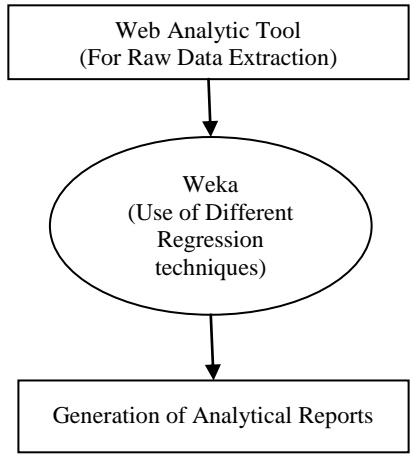

**Fig 1.** Use of Weka in this Experiment

Effort based on Cost: Company spends on Banner Ad Spend and PPC to attract visitors to their website and increase page views.

Improving Page Views has been identified as the major goal for the website for which they are investing in the above efforts.

While working on this data set,it was identified that Weka is powerful enough to work on small data set as well. When the data is large,different algorithms used in mining has many instances of different variations so they have better options to learn and then give desired results .This makes these algorithms to predict the findings close to the actual finding and thus making them very effective in turn. But the real problem is with small data set. Not much variation among the data is found in the small data sets which makes it difficult for any algorithm to learn and predict future movement efficiently.

## 3. EXPERIMENT

### 3.1 Data Set Detail

Data set [Table 1] used in this experiment is from a mid size company and rages from April 2011 to June 2012 having 7 attributes namely,

1. Subscriber (Total)-ST
2. Banner Ad Spend-BAS
3. PPC spend-PPC
4. Reminder E mails sent –REM
5. Videos Upload -VU
6. Page view-PV
7. Month

So the data set have a total of 7 attributes and 15 instances of each attribute. Company considers Page view as its assets and wants to find out its relationships with other attributes as shown in Fig 2.

Main objective is to find out the attributes directly responsible for the page views. is to find the optimal solution using different techniques of mining using Weka to find the best relationship for the company. Our aim is also to find the power of Weka on small data set. How accurately it can predict the future movements while given a narrow chance to learn from data.

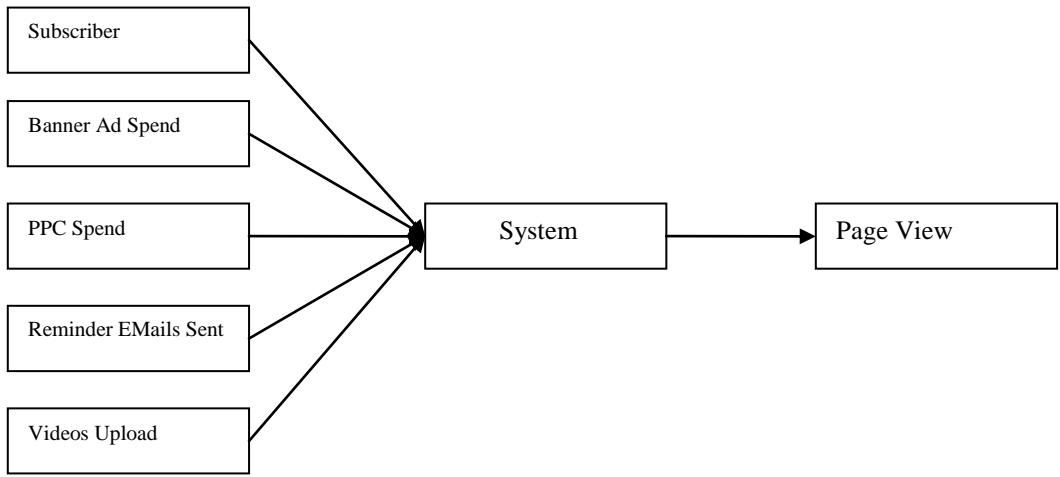

**Fig. 2** Company Input Output Details

**Table 2. Analysis Table**

| Ori PV | As per Fig 3c Pre./ Diff. Err% | As per Fig 3e Pre./ Diff. Err% | As per Fig 3g Pre./ Diff. Err% | As per Fig 3d Pre./ Diff. Err% | As per Fig 3f Pre./ Diff. Err% |
|---|---|---|---|---|---|
| 20k | 82075 / 62,075 / 310.375 | 141951 / 121,951 / 609.755 | 54318 / -34,317 / -172 | 51677 / -31,676 / -159 | 34,053 / -14,053 / -71 |
| 100k | 122368 / 22,367 / 22.368 | 139223 / 39,223 / 39.223 | 97317 / 2,684 / 3 | 100028 / -28 / -1 | 131,724 / -31,724 / -32 |
| 200k | 172733 / -27,267 / -13.634 | 187656 / -12,345 / -6.173 | 150549 / 49,452 / 25 | 156728 / 43,273 / 22 | 200,000 / 0 / -1 |
| 250k | 258909 / 8,909 / 3.564 | 259068 / 9,068 / 3.627 | 260284 / -10,284 / -5 | 261642 / -11,641 / -5 | 280,052 / -30,052 / -13 |
| 300k | 291308 / -8,693 / -2.898 | 291865 / -8,136 / -2.712 | 258030 / 41,971 / 14 | 259821 / 40,179 / 14 | 285,896 / 14,104 / 5 |
| 400k | 405251 / 5,251 / 1.313 | 403649 / 3,649 / 0.913 | 409121 / -9,120 / -3 | 399979 / 22 / 1 | 390,942 / 9,058 / 3 |
| 550k | 546349 / -3,651 / -0.664 | 543034 / -6,967 / -1.267 | 591375 / -41,375 / -8 | 577329 / -27,328 / -5 | 550,000 / 0 / -1 |
| 650k | 657569 / 7,568 / 1.165 | 623342 / -26,659 / -4.102 | 667621 / -17,620 / -3 | 660063 / -10,062 / -2 | 662,888 / -12,888 / -2 |
| 700k | 678259 / -21,741 / -3.106 | 653772 / -46,229 / -6.605 | 699142 / 859 / 1 | 699389 / 612 / 1 | 700,000 / 0 / 1 |
| 800k | 743938 / -56,063 / -7.008 | 741948 / -58,053 / -7.257 | 785714 / 14,287 / 2 | 782411 / 17,590 / 3 | 768,048 / 31,952 / 4 |
| 900k | 858830 / -41,171 / -4.575 | 846335 / -53,666 / -5.963 | 819846 / 80,155 / 9 | 817285 / 82,716 / 10 | 831,864 / 68,136 / 8 |
| 1150k | 1098891 / -51,110 / -4.445 | 1066700 / -83,301 / -7.244 | 1144763 / 5,237 / 1 | 1137599 / 12,402 / 2 | 1,150,000 / 0 / 1 |
| 1300k | 1360869 / 60,869 / 4.683 | 1297356 / -2,645 / -0.204 | 1334870 / -34,869 / -3 | 1385749 / -85,748 / -7 | 1,436,100 / -136,100 / -11 |
| 1475k | 1544234 / 69,233 / 4.694 | 1556915 / 81,915 / 5.554 | 1474142 / 859 / 1 | 1475022 / -22 / -1 | 1,475,000 / 0 / -1 |
| 1900k | 1873442 / -26,559 | 1942216 / 42,216 | 1900859 / -859 | 1901074 / -1,074 | 1,900,000 / 0 |

| | -1.398 | 2.222 | -1 | -1 | 1 |
|---|---|---|---|---|---|

## 3.2 Problem Statement

Endeavour is to answer following queries that are generated from previous discussions.
1. Is there any /direct or indirect relationship between the attributes?
2. What are the basic inputs that contribute in achieving output?
3. What are the mutual dependencies between inputs that can impact output?
4. What are the different areas where company needs to focus ?
5. Company is gaining profit or Loss?

## 3.3 Experimental Setup

This Experiment involves four different tools namely Web analytic tool to collect the data, Notepad++[8] for data conversion to ARFF format, Weka-the data mining tool and MS Excel for some extra mining of data.

This experiment follows the different steps as described in Fig. 1.
Step-1 Data was collected from WWW using Web analytical tool.
Step-2 Redundant data was removed.
Step-3 Data was converted to ARFF file Format.(Attribute Relationship File Format).
Step-4 (a) Multiple Linear Regression technique was executed keeping PV as dependent variable while ST, PPC, BAS, REM, VU as independent Variables.
Step-4 (b) Multiple Linear Regression techniques was again applied keeping PV as dependent variable while ST,PPC,BAS,VU as independent variables.
Step-4 c) SMO Regression technique with normalization was applied while keeping dependent and independent variable as in Step4(a).
Step-4 (d) SMO Regression technique with Standardization was applied while keeping dependent and independent variables as in Step 4(a).
Step-4 (e) SMO Regression technique without any normalization and without any standardization was applied while keeping dependent and independent variables as in step4 (a).
Step-5 Various output were consolidated in Fig 3.
Step-6 Table No 3 having Original Page Views,Predicted Page views, Difference between Page Views and Error Percentage was created for processing.
Step-7 Table No 8 having total cost and expected cost with respect to PPC spend and Banner Ad spend and profit was generated using MS Excel.

## 4. RESULT ANALYSIS

Results obtained are consolidated in Fig 3 as Appendix 1 and its analysis is consolidated in Table 2.From Careful examination of the results ,authors try to answer the questions raised in above paragraph.

## 4.1 Mining data relationships using WEKA

The value of Correlation coefficient tells us that different attributes of the data set are closely related to each other in one to one relationship. Relationship among different attributes are described in Fig 3a in the form of correlation. Based on the information obtained from fig3a, it can be determined that there are only two attributes namely subscribers total and reminder email Sent which are showing a higher degree of correlation with Page view. Inference: Reminder emails sent to the subscribers is the most important ways to increase the page views.

## 4.2 Finding Most Important Growth factor

From above discussion and from the different figures of fig 3 ,it can be determined that the most important growth factor for the company are basically two .One is the number of subscribers and the other is the Reminder E mails sent.

Based on Fig 3d,3f,3g it can also be predicted that the third most important factor is PPC spend although its weightage is much lesser than the earlier two factors. Other variable like Banner ad Spend and Videos upload has very little significance on Page Views.

Inference: Two most important and necessary growth factors are Reminder Email and Subscribers total.

## 4.3 Dependencies between input and output.

Page view is found to be having one to many relationships prominently with Subscribers and Reminder E mails Sent. Its having very weak relationship with Banner Ad Spend, PPC spend and videos upload.

## 4.4 Areas needs to be focussed.

Further from Table 3,it can be determined that the Banner Ad Spend and PPC spend are the main area where the company is investing both in terms of money as well as resources. Although the 5$^{th}$ column is showing the profit area but a closer look at column 6 tells us actual finding about profit. It tells us that out of 14 months Profit shows an increase in growth only for 8 times while it showed a decline in profit for 6 times which is not appositive sign keeping in mind that cost was calculated keeping only two attribute namely Banner Ad spend and PPC spend in mind and both of these attributes are of less significant importance according to Fig 3.

Inference: From above paragraph it can be determined that the company seriously needs to either preempt itself on spending on these two attributes or need to focus more on the way the spending is done using these attributes.

## 4.5 About Profit /Loss

It can be concluded from table 4 that the profit percentage, though fluctuating, never reached negative mark. Also Table 3 stats that most of the time Page view reaches close to the predicted value. These predicted values were heavily based on Reminder e mails Sent and subscribers total. These two attributes are not responsible for utilizing company resources in terms of money.

Inference: Company is making profit but to make it more worthy,it needs to concentrate more on strategies focusing Banner Ad Spend and PPC spend.

Table 3. Company Growth Table

| PV | Total cost | Estimated Cost | DIFF | Pro % | INC (+) /DEC (-) |
|---|---|---|---|---|---|
| 20k | $650 | 20000 | 0 | 0 | NIL |
| 100k | $2,500 | 76924 | 23,076 | 24 | INC |
| 200k | $3,000 | 92308 | 107,692 | 54 | INC |
| 250k | $1,600 | 49231 | 200,769 | 81 | INC |
| 300k | $1,550 | 47693 | 252,307 | 85 | INC |
| 400k | $9,734 | 299508 | 100,492 | 26 | DEC |
| 550k | $16,832 | 517908 | 32,092 | 6 | DEC |
| 650k | $11,339 | 348893 | 301,107 | 47 | INC |
| 700k | $3,489 | 107354 | 592,646 | 85 | INC |
| 800k | $11,723 | 360708 | 439,292 | 55 | DEC |
| 900k | $14,372 | 442216 | 457,784 | 51 | DEC |
| 1,150k | $29,282 | 900985 | 249,015 | 22 | DEC |
| 1,300k | $29,378 | 903939 | 396,061 | 31 | INC |
| 1,475k | $24,783 | 762554 | 712,446 | 49 | INC |
| 1,900k | $55,486 | 1707262 | 192,738 | 11 | DEC |

## 5. CONCLUSION AND FUTURE SCOPE

This experiment was done on real time web data and was a sincere attempt to answer some questions from a company growth perspective.

Out of basic three data mining technique, namely classification, clustering and regression, regression technique was used because of the nature of the data. Other technique were, though powerful, but was not suitable to mine information from this data.

One very important challenge this analysis pose to the researcher is to find a suitable mathematical model to mine the information for comparison purposes as most of the data mining tools can mine the data and gives you various result but a human interpretation is very much necessary to retrieve necessary information. A mathematical model will be very beneficial to compare the results obtained using various techniques and to choose the best one.

**APPENDIX 1**

|     | ST   | BAS  | PPC  | REM  | VU   | PV   |
|-----|------|------|------|------|------|------|
| ST  | 1    | 0.95 | 0.9  | 0.93 | 0.78 | 0.99 |
| BAS | 0.95 | 1    | 0.98 | 0.84 | 0.82 | 0.94 |
| PPC | 0.9  | 0.98 | 1    | 0.74 | 0.78 | 0.88 |
| REM | 0.93 | 0.84 | 0.74 | 1    | 0.75 | 0.95 |
| VU  | 0.78 | 0.82 | 0.78 | 0.75 | 1    | 0.77 |
| PV  | 0.99 | 0.94 | 0.88 | 0.95 | 0.77 | 1    |

**Fig. 3a Correlation Matrix among attributes**

| Sr. No. | Scheme name | Correlation Coefficient |
|---------|-------------|------------------------|
| 1.      | Linear Regression(Fig 3c) | 0.9973 |
| 2.      | Linear Regression(Fig 3e) | 0.9931 |
| 3.      | SMO Regression(Fig 3g)    | 0.9975 |
| 4.      | SMO Regresion(Fig 3d)     | 0.9977 |
| 5       | SMO Regresion(Fig 3f)     | 0.997  |

**Fig. 3b Consolidated CC as per scheme**

Page_Views =

    10.0731 * Subscribers_total +
    68.0727 * Remindr_Emails_Sent +
    72001.7239

**Fig 3c Linear Regression Result**

weights  (not support vectors):

    + 0.5108  *  (normalized) Subscribers_total
    + 0.0752  *  (normalized) Banner_Ad_Spent
    + 0.1439  *  (normalized) PPC_Spend
    + 0.3368  *  (normalized) Reminder_Emails_sent
    - 0.0676  *  (normalized) Videos_Upload
    + 0.0315

**Fig 3d SMOreg with Normalization**

Page_Views =

    12.5469 * Subscribers_total +
    14.8024 * Banner_Ad_Spend +
    -28.6031 * PPC_Spend +
    147995.9826

**Fig 3e Linear Regression result with limited attribute**

weights (not support vectors):

    + 7.5597   *  Subcribers_Total
    - 14.6798  *  Banner_Ad_Spend
    + 40.8236  *  PPC_Spend
    + 123.227  *  Reminder_Emails_sent
    - 197.3539 *  Videos_Upload
    + 19693.6481

**Fig 3f SMOReg Results**

weights (Not Support vectors) :

    + 0.5531  *  (standardized) Subscriber_total
    + 0.0306  *  (standardized) Banner_Ad_spend
    + 0.1491  *  (standardized) PPC_Spend
    + 0.378   *  (standardized) Remoinder_Emails_sent
    - 0.08    *  (standardized) Videos_Upload
    - 0.0036

**Fig 3g SMOReg with Standardization**

Attribute  Subset  Evaluator(Supervised,Class(numeric)
:6 Page_views
        CFS Subset Evaluators
        Including locally predictive attributes

Selected Attributes: 1,4:2
                Subscriber_total
                Reminder_Emails_Sent

**Fig 3h Evaluation of Best Fit Attributes**

**Fig 3 Different Results obteained using Weka**